\documentclass[aps,prl,twocolumn,superscriptaddress]{revtex4}
\usepackage{amsmath}
\usepackage{amssymb}

\usepackage{amsmath,amssymb}
\usepackage[usenames]{color}
\usepackage{mathbbol}
\usepackage{amssymb}
\usepackage{grffile}
\usepackage[pdftex]{graphicx}
\usepackage{amsmath, amstext, amssymb, amsfonts, amsxtra}
\usepackage{textcomp}
\usepackage{xspace} 
\usepackage[colorlinks]{hyperref}

\DeclareSymbolFontAlphabet{\amsmathbb}{AMSb}

\newcommand{\tbf}[1]{\textbf{#1}}
\newcommand{\abs}[1]{|#1|}

\newcommand{\bonnpi}{Physikalisches Institut, University of Bonn, Nussallee 12, 53115 Bonn, Germany}

\begin{document}

\title{Slow interaction quench in BCS superconductors: emergence of pre-formed pairs}

\author{Johannes Kombe}
\affiliation{\bonnpi} 
\author{Jean-S\'ebastien Bernier}
\affiliation{\bonnpi}
\author{Michael K\"ohl}
\affiliation{\bonnpi}
\author{Corinna Kollath}
\affiliation{\bonnpi} 

\begin{abstract}
  We investigate the non-equilibrium behavior of BCS superconductors subjected to slow ramps of
  their internal interaction strength. We identify three dynamical regimes as a function of ramp duration.
  For short ramp times, these systems become non-superconducting; however, fermions with opposite momenta
  remain paired albeit with reduced amplitudes, and the associated pair amplitude distribution is non-thermal.
  In this first regime, the disappearance of superconductivity is due to the loss of phase coherence between pairs.
  By contrast, for intermediate ramp times, superconductivity survives but the magnitude of the order parameter
  is reduced and presents long-lived oscillations. 
  Finally, for long ramp times, phase coherence is almost fully retained during the slow interaction quench,
  and the steady-state is characterized by a thermal-like pair amplitude distribution.
  Using this approach, one can therefore dynamically tune the coherence between pairs in order to control
  the magnitude of the superconducting order parameter and even
  engineer a non-equilibrium state made of pre-formed pairs.
\end{abstract}

\date{\today}

\maketitle

%
%
The properties of quantum materials are extremely sensitive to external stimuli.
In these systems, the interactions associated with, for example, the spin, charge,
lattice and orbital degrees of freedom are often similar in magnitude with
the electronic kinetic energy. The delicate balance between competing states
can therefore be readily altered via external perturbations leading to the emergence of
novel properties. Taking advantage of this distinctive characteristic of
quantum materials and building on the tremendous technical progress achieved in the
last decade, scientists can now dynamically engineer complex states and follow their
non-equilibrium evolution. Phase transitions were photo-induced in strongly
interacting solid state compounds using ultrafast optical
pulses~\cite{BasovDressel2011, Orenstein2012, ZhangAveritt2014, GiannettiMihailovic2016},
and similar achievements were also reported in ultracold atomic systems using time-dependent
electromagnetic fields~\cite{BlochZwerger2008, PolkovnikovVengalattore2011}.
For example, in a striped-order cuprate, a Josephson plasmon, a hallmark of the superconducting
state, was activated above the critical temperature by the application of midinfrared
femtosecond pulses~\cite{FaustiCavalleri2011}. While these results are truly remarkable,
the mechanisms underlying the non-equilibrium dynamics of strongly correlated matter
are still being investigated.

Identifying the processes governing the evolution of order parameters when interactions
are tuned over time remains an open question. As order parameters are
global quantities often made up from sums of local or quasi-local
(in position or momentum space) expectation values, one would like to understand how the
time-dependent behavior of these different local components conspires to control
the dynamics of the global order parameter. 
Superconductors described by the Bardeen-Cooper-Schrieffer (BCS) theory of superconductivity
constitute an interesting example as in these systems the collective order parameter is built
out of individual Cooper pair states labeled by their internal momentum. A similar situation
also arises in magnets where the magnetization is a sum over all local 
spins.

Going back to our first example, superconductivity, in this case
understanding the dynamics arising from the subtle
interplay between the BCS collective mode and its constituting elements
following a sudden quench of the pairing strength has been the focus of
various works~\cite{BarankovSpivak2004, WarnerLeggett2005, SimonsBurnett2005, YuzbashyanAltshuler2006, BarankovLevitov2006, YuzbashyanDzero2006, PapenkortKuhn2007,
  DzeroAltshuler2007, Gurarie2009, DzeroAltshuler2009, Galitski2010, ScottStringari2012, YuzbashyanFoster2015}.
The renewed interest for this problem has been triggered in part by the
possibility in dilute fermionic gases cooled below degeneracy to control the interaction strength using
Feshbach resonances~\cite{BlochZwerger2008}. Using both analytical and numerical methods, three different dynamical
regimes were unveiled in a space spanned by the ratio of the equilibrium superconducting order parameters of the initial state
to the one of the final state. When this ratio is sufficiently small, the order parameter
oscillates without damping, while for intermediate ratios, it is damped, exhibits decaying oscillations
and saturates at an asymptotic value. In contrast, for larger ratios, the order parameter is overdamped
following the sudden interaction quench and ultimately superconductivity is destroyed.

\begin{figure} 
\includegraphics[width=1\columnwidth]{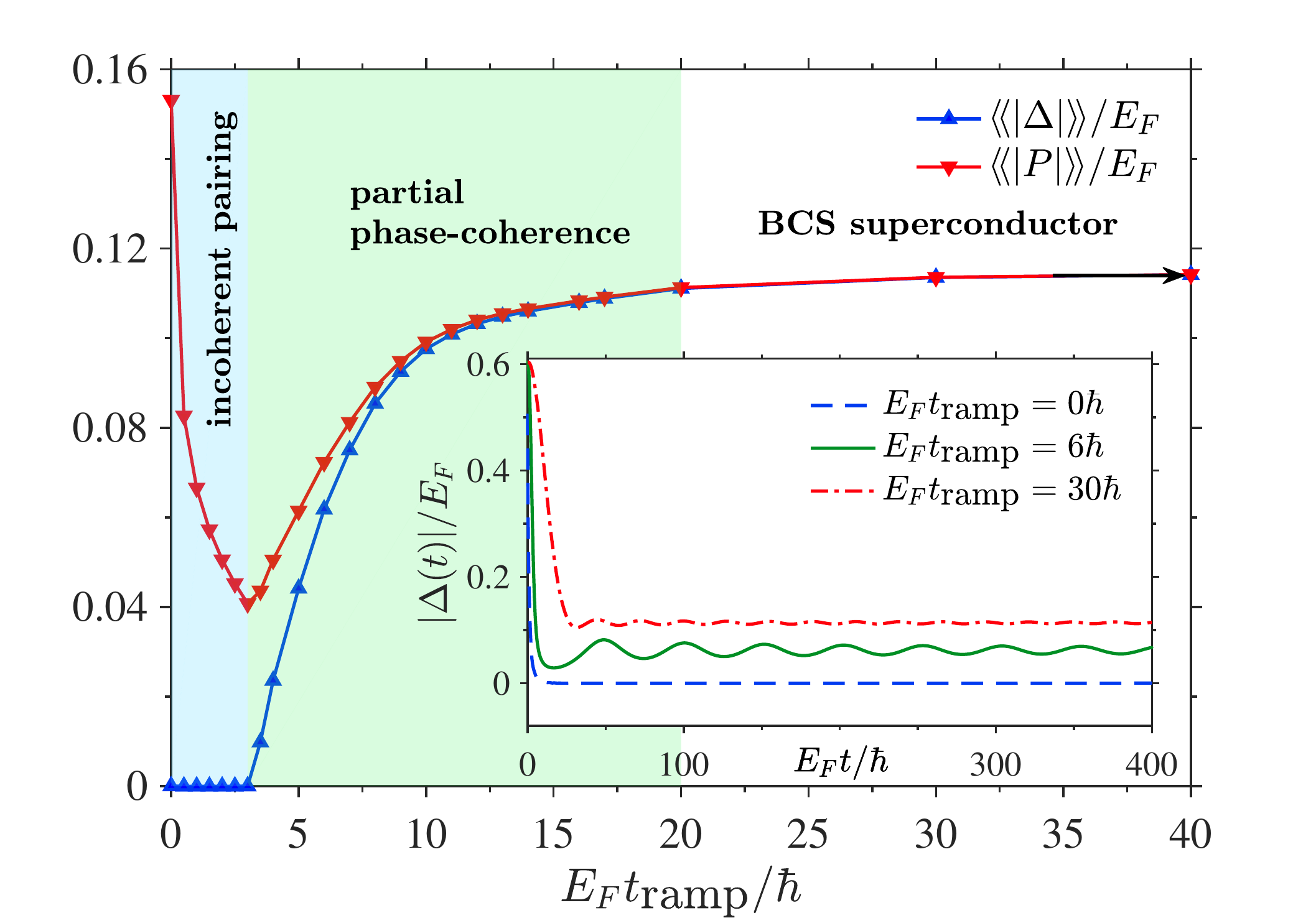}
\caption{(color online) \label{fig:OPvsTime} 
  Time-averaged value of the superconducting order parameter, $\langle \! \langle |\Delta| \rangle \! \rangle$,
  and of the sum over the magnitude of the pair amplitudes, $\langle \! \langle |P| \rangle \! \rangle$,
  as a function of ramp duration. The interaction strength is ramped down from
  $1/(k_\text{F}a) = -0.1072$ ($|\Delta_{0,\text{i}}| = 0.60 E_\text{F}$)
  to $1/(k_\text{F}a) = -1.3493$ ($|\Delta_{0,\text{f}}| = 0.11 E_\text{F}$), and the time-average is taken between $100 \hbar/E_\text{F}$
  and $400 \hbar/E_\text{F}$. These two quantities signal three different dynamical regimes. For short quenches,
  the system is characterized by pre-formed pairs (incoherent pairing state). For intermediate quench durations,
  superconductivity is maintained but with only partial phase-coherence; while, for longer ramp times, phase coherence
  is mostly unaffected and the order parameter asymptotes to $\Delta_{0,\text{f}}$ (value marked by the arrow).
  Inset: evolution of the order parameter as a function of time for three different ramp durations.
}
\end{figure}

Experimentally, quenches are typically realized within a finite ramp time. This can be achieved, for example, by exciting phononic modes in
solids~\cite{SentefKollath2016} or in cold gases by ramping a magnetic field. We study in this article the dynamics of the BCS order parameter
when the interaction between fermions is slowly changed in time. Focusing on the dynamical region where a sudden change of the interaction
strength would obliterate the superconducting order parameter, we find that when the same interaction change is carried out at a slower rate
three different regimes emerge.
As shown in Fig.~\ref{fig:OPvsTime}, for sufficiently
short ramp times, the system becomes non-superconducting as the interaction strength is lowered to
its new value.
However, in contrast to the usual non-superconduting thermal state above the critical temperature,
Cooper pairs are still present in this dynamically generated non-superconducting state,
but with reduced amplitudes.
This is signalled by the finite value taken by the sum over the magnitude of the momentum-dependent pair
amplitudes. Hence, the loss of phase coherence between these pairs is responsible here for the disappearance of
superconductivity.
Consequently, for short ramp times,
dynamically engineering a state made of incoherent pre-formed
pairs, the so-called phase disordered superconductor~\cite{EmeryKivelson1995}, is possible.
With increasing ramp time, partial phase coherence is restored and the superconducting order parameter acquires a
finite value. Whereas in the incoherent pairing regime, the sum over the magnitude of pair
amplitudes decreases, this quantity increases again once superconductivity is re-established in the second
dynamical regime displaying a surprising non-monotonic behavior.
Finally, for long ramp times, full phase coherence is almost retained during the slow
interaction change, and the amplitude distribution of the individual pairs becomes thermal-like. In the remaining of this article,
we present in more details the subtle mechanism responsible for the dynamics of the superconducting order parameter
during and after a slow ramp down of the interaction strength.

%
%
To study this dynamics, we consider a situation applicable to both
solid state systems and cold atom gases: a three-dimensional gas made of two species of fermions described
by the BCS $s$-wave Hamiltonian
\begin{align} \label{eq:H_BCS}
  H_{\text{BCS}} &= \!\!\! \sum_{\tbf{k}, \sigma = \{1,2\}} \!\!\! \epsilon_{\tbf{k}}~n_{\tbf{k}, \sigma} +
  \sum_{\tbf{k}} \Big[ \Delta~c^{\dagger}_{\tbf{k}, 1} c^{\dagger}_{-\tbf{k}, 2} + \text{h.c.} \Big],
\end{align}
where $c_{\tbf{k},\sigma}^{(\dag)}$ are the fermionic annihilation (creation) operators, $n_{\tbf{k},\sigma}$
is the particle number operator of momentum $\tbf{k}$ and species $\sigma=\{1, 2 \}$, and
$\epsilon_{\tbf{k}} = \hbar^2\tbf{k}^2/(2m)$ is the single-particle dispersion. 
The superconducting order parameter enters this Hamiltonian as
\begin{align} \label{eq:OP}
  \Delta &= \frac{g}{V} \sum_{\tbf{k}} P_{\tbf{k}}
\end{align}
with $V$ the system volume and $g$ the interaction strength. Here, the expectation value 
$P_{\tbf{k}} = \langle c_{-\tbf{k}, 2} c_{\tbf{k}, 1} \rangle$ relates to individual Cooper pairs.
Additionally, to assess the individual pairing strength, we introduce a second quantity corresponding to the
sum over the magnitude of the momentum-dependent pair amplitudes
\begin{align}
P &= \frac{g}{V} \sum_{\tbf{k}} \abs{P_\tbf{k}}.
\end{align}
For ultracold gases, the strength of the interaction between the fermions of two different hyperfine states
can be tuned via Feshbach resonances~\cite{BlochZwerger2008}, and at sufficiently low temperatures the 
$s$-wave scattering is the dominant contribution. In this situation, the interaction can
be parametrized by a single parameter, the $s$-wave scattering length $a$, via
$\frac{1}{k_\text{F}a} = \frac{8 \pi E_\text{F}}{g k_\text{F}^{3}} + \frac{2}{\pi}\sqrt{\frac{E_\text{C}}{E_\text{F}}}$ with $k_\text{F}$ and $E_\text{F}$
the Fermi momentum and energy, respectively, and $E_\text{C}$ a suitably chosen energy cutoff.

The equilibrium phase diagram for a system described by this Hamiltonian has been thoroughly
studied~(see \cite{ZwierleinKetterle2008} and references therein).
For $\frac{1}{k_{F}a} < 0$, the interaction is attractive and below the critical temperature, $T_c$,
this system arranges into a superfluid of Cooper pairs. In this situation, the value of the
superconducting order parameter decreases with increasing temperature as the thermal generation of single-particle
excitations leads to the breaking of Cooper pairs.

Here, we consider a slower interaction change of the parameter $\frac{1}{k_{F}a} < 0$ using the schedule
\begin{align} \label{eq:Schedule}
  \frac{1}{k_{F}a(t)}  = \frac{1}{k_{F}a(t_\text{f})} + h(t, t_\text{i}, t_\text{f})
  \Bigg[1 -  \sin^{2}\Big(\frac{\pi}{2} \frac{t - t_\text{i}}{t_{\text{ramp}}}\Big)\Bigg], \nonumber
\end{align}
where $h(t, t_\text{i}, t_\text{f}) = \Theta(t_\text{f} - t) \Big[ 1/(k_{F}a(t_\text{i})) - 1/(k_{F}a(t_\text{f})) \Big]$,
$t_{\text{ramp}} = t_\text{f} - t_\text{i}$ with $t_\text{i}$ and $t_\text{f}$ the times at which the interaction ramp
begins and ends, and $\Theta$ is the Heaviside function.
$t_{\text{ramp}} \rightarrow 0$ corresponds to a sudden interaction change while $t_{\text{ramp}} \rightarrow \infty$ would
correspond to an adiabatic interaction change. We focus on the situation where the interaction strength is ramped down such that the
initial equilibrium value of the superconducting order parameter, $\Delta_{0,\text{i}}$, is larger than the final
equilibrium value, $\Delta_{0,\text{f}}$.

To understand the dynamics of the order parameter, we obtain
a set of coupled differential equations connecting the superconducting order parameter to the
expectation values of individual pairs and atom densities:
\begin{eqnarray}
  \hbar \frac{\partial}{\partial t} \langle c_{-\tbf{k},2} c_{\tbf{k},1} \rangle &=&
  i  \{-2\epsilon_{\tbf{k}} \langle c_{-\tbf{k},2} c_{\tbf{k},1} \rangle \nonumber \\
  && + \Delta (\langle n_{\tbf{k},1} \rangle + \langle n_{-\tbf{k},2} \rangle - 1)  \},  \nonumber \\
  \hbar \frac{\partial}{\partial t} \langle n_{\tbf{k},1} \rangle & = &  - 2~\text{Im}\{\Delta^{*}  \langle c_{-\tbf{k},2} c_{\tbf{k},1} \rangle \},  \nonumber \\
  \hbar \frac{\partial}{\partial t} \langle n_{-\tbf{k},2} \rangle & = & - 2~\text{Im}\{\Delta^{*}  \langle c_{-\tbf{k},2} c_{\tbf{k},1} \rangle \}.
\end{eqnarray}
Solving numerically this set of equations together with the self-consistency condition for $\Delta$, Eq.~\ref{eq:OP},
we compare and contrast the non-equilibrium evolution due to a sudden
interaction change, $t_{\text{ramp}} \rightarrow 0$, to ones due to longer ramp times.

%
%
The main results are summarized in Fig.~\ref{fig:OPvsTime} comparing the time-averaged
superconducting order parameter, $\langle \! \langle |\Delta| \rangle \! \rangle$, and
the time-averaged sum over the magnitude of the pair amplitudes, $\langle \! \langle |P| \rangle \! \rangle$.
For a sudden interaction change and for ramp times up to $3\hbar/E_\text{F}$, the superconducting order
parameter is found to average to zero whereas, for slower interaction ramps, this order parameter retains a finite value.
The precise ramp duration at which the crossover occurs depends on the interaction strength and the cutoff.

In contrast, we find that the sum over the magnitude of the
pair amplitudes, remains finite for all ramp times.
This important difference in behavior between these two quantities signals that the evolution of the
relative phase between individual pairs plays a crucial role in the dynamics. As the amplitude of pairs is
reduced but remains finite, the destruction of superconductivity for fast ramps is
associated with the loss of phase coherence between pairs. Therefore, phase unlocking is the main mechanism responsible for the
suppression of superconductivity. This is in stark contrast to the finite-temperature equilibrium scenario where
superconductivity is suppressed by an increase in thermal fluctuations resulting
in pair breaking. Interestingly, this result implies that stabilizing a state made of pre-formed
pairs is possible via a fast ramp. Within the scope of the BCS model, this state is long-lived; however, in real
materials the presence of various coupling mechanisms could likely affect the long-time stability of this state.

Unexpectedly, $\langle \! \langle |P| \rangle \! \rangle$, the sum over the magnitude of the pair amplitudes,
is non-monotonous as a function of ramp time:
it first decreases with increasing ramp times and then increases again. For a sudden quench, this quantity
has a large value due to the freezing of the initial state which
is then projected onto the new Cooper pairs. Within this new basis, this frozen state contains excited
quasiparticle pairs which contribute to the sum over the magnitude of pair amplitudes.
As these quasiparticle pairs are not coherent, their contributions to the total order parameter dephase
after a short time resulting in the suppression of the superconducting order parameter. For short but finite ramp times, the
same mechanism persists until the sum over the magnitude of the pair amplitudes reaches a minimum.  For longer ramp times,
this quantity rises again and the dephasing becomes less important such that the value of the superconducting order parameter
is finite at longer times. 

\begin{figure} 
\includegraphics[width=1\columnwidth]{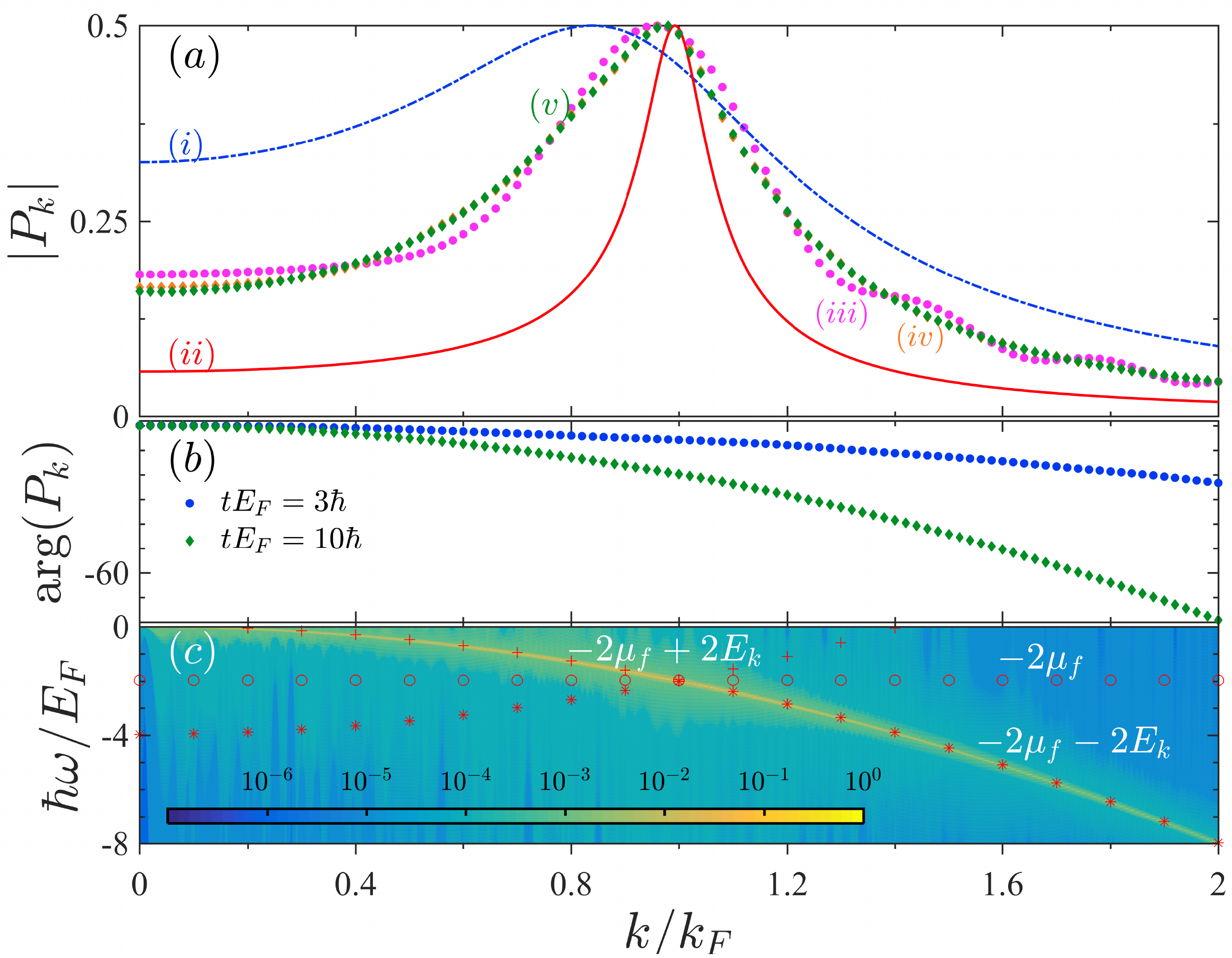}
\caption{(color online) \label{fig:PanelsTramp0} Sudden quench of the interaction strength from
  $1/(k_\text{F}a) = -0.1072$ to $1/(k_\text{F}a) = -1.3493$. (a) Distribution of the magnitude of the pair amplitude
  as a function of momentum: (i, blue) ground state at $1/(k_\text{F}a) = -0.1072$; (ii, red) ground state
  at $1/(k_\text{F}a) = -1.3493$; (iii, pink) snapshot at $3\hbar/E_\text{F}$; (iv, orange) snapshot at $10\hbar/E_\text{F}$;
  and (v, green) time-average between $100\hbar/E_\text{F}$ and $400\hbar/E_\text{F}$. This distribution is already
  hardly distinguishable form its steady-state configuration at $10\hbar/E_\text{F}$. It is non-thermal and signals the presence of
  pre-formed pairs. (b) Phase of the pair amplitude as a function of momentum. Rapid phase unlocking is responsible for the 
  destruction of superconductivity. (c) Fourier transform of the momentum-dependent pair amplitude $|\mathcal{F}[\text{Im}[P_k(t)]]|$.
  The sudden quench generates quasiparticle pair excitations along the parabolic line $\pm 2 E_\tbf{k} - 2\mu_\text{f}$
  (marked by the red crosses and stars respectively; red circles mark the coherent evolution at $-2\mu_\text{f}$).}
\end{figure}

\begin{figure} 
\includegraphics[width=1\columnwidth]{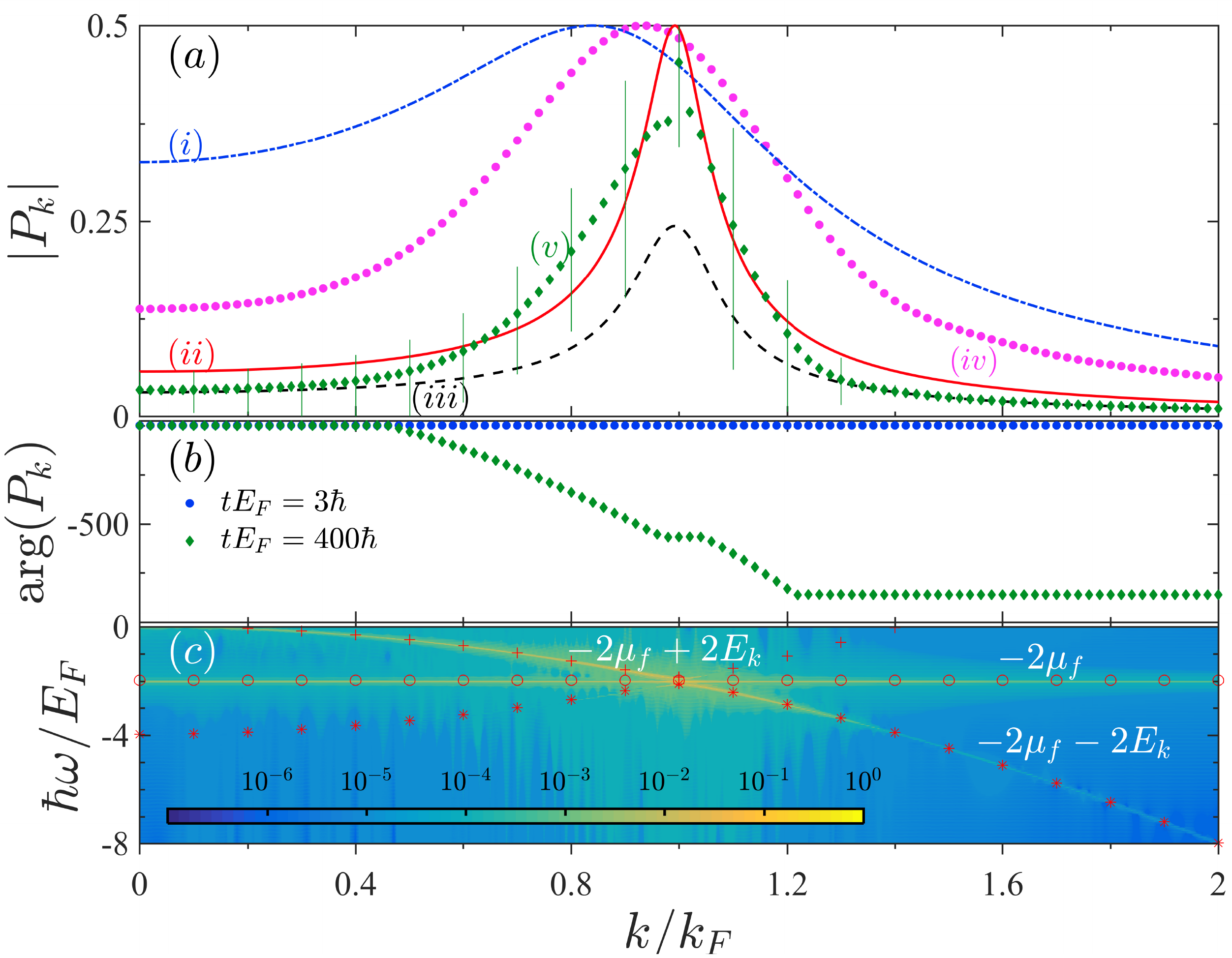}
\caption{(color online) \label{fig:PanelsTramp6} Slow quench performed in $E_\text{F} t_\text{ramp} = 6 \hbar$ (same
  interaction strengths as in Fig.~\ref{fig:PanelsTramp0}).
  (a) Distribution of the magnitude of the pair amplitude as a function of momentum: (i, blue) and (ii, red)
  same as in Fig.~\ref{fig:PanelsTramp0}; (iii, black) thermal distribution at $1/(k_\text{F}a) = -1.3493$,
  $T = 0.89 T_c$ is chosen such that $\langle \! \langle |\Delta| \rangle \! \rangle$ = $\Delta(T)$;
  (iv, pink) snapshot at $3\hbar/E_\text{F}$; (v, green) time-average between $100\hbar/E_\text{F}$ and $400\hbar/E_\text{F}$.
  The time-averaged distribution is non-thermal, and $|P_k|$ exhibit strong oscillations
  represented, together with (v), by vertical bars (peak-to-peak amplitude of the oscillations). (b) Phase of the pair amplitude
  as a function of momentum.
  (c) Fourier transform of the momentum-dependent pair amplitude $|\mathcal{F}[\text{Im}[P_\tbf{k}(t)]]|$.
  While this quench generates quasiparticle pair excitations, all $P_\tbf{k}$ signals have a strong in-phase component
  at $-2\mu_\text{f}$.}
\end{figure}

\begin{figure} 
\includegraphics[width=1\columnwidth]{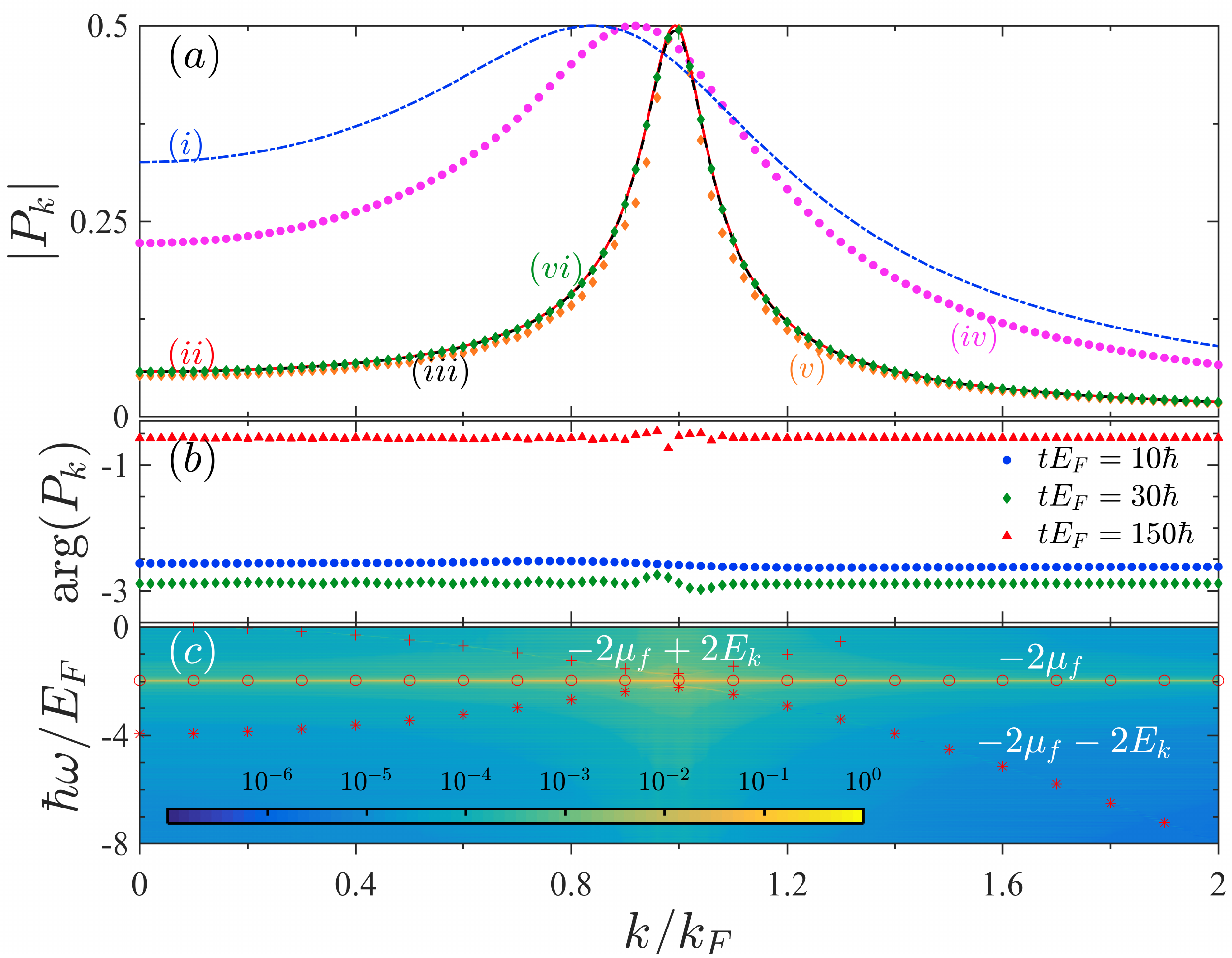}
\caption{(color online) \label{fig:PanelsTramp30} Slow quench performed in $E_\text{F} t_\text{ramp} = 30 \hbar$ (same
  interaction strengths as in Fig.~\ref{fig:PanelsTramp0}).
  (a) Distribution of the magnitude of the pair amplitude as a function of momentum: (i, blue) and (ii, red)
  same as in Fig.~\ref{fig:PanelsTramp0}; (iii, black) thermal distribution at $1/(k_\text{F}a) = -1.3493$,
  $T = 0.35 T_c$ is chosen such that $\langle \! \langle |\Delta| \rangle \! \rangle$ = $\Delta(T)$;
  (iv, pink) snapshot at $10\hbar/E_\text{F}$; (v, orange) snapshot at $30\hbar/E_\text{F}$; (vi, green)
  time-average between $100\hbar/E_\text{F}$ and $400\hbar/E_\text{F}$. The steady-state distribution
  is thermal. (b) Phase of the pair amplitude as a function of momentum. Phase
  coherence is only slightly lost near $k_\text{F}$. (c) Fourier transform of the momentum-dependent pair
  amplitude $|\mathcal{F}[\text{Im}[P_\tbf{k}(t)]]|$. Quasiparticle pairs are only generated near $k_\text{F}$ (red crosses and stars).
  $P_k$ are dominated by the coherent phase evolution at $-2\mu_\text{f}$.}
\end{figure} 

In the following, we analyze carefully the behavior of the excitations responsible for the emergence of
the non-equilibrium states detailed above. Useful information can be obtained by analyzing
the momentum distribution of the Cooper pair amplitudes. As displayed in Fig.~\ref{fig:PanelsTramp0},
initially the distribution of pair amplitudes follows the zero temperature and
interaction-dependent expression $P_{\tbf{k}} = \frac{1}{2} \sqrt{1 - \xi_{\tbf{k}}^2/E_{\tbf{k}}^2}$ with $\xi_{\tbf{k}} = \epsilon_{\tbf{k}}-\mu_\text{i}$
where $\mu_\text{i}$ is the chemical potential at the initial interaction strength. For $\frac{1}{k_\text{F} a} = -0.1072$,
this distribution has a maximum close to the Fermi momentum and drops down for larger momentum values.
In contrast, the distribution corresponding to the final ground state of an adiabatic quench to $\frac{1}{k_\text{F} a} = -1.3493$
is strongly peaked around the Fermi momentum and is much lower in value than the intial distribution. 

To understand the time evolution of the distribution of pair amplitudes, we consider both snapshots of $|P_k|$ at
particular times and values denoted by $\langle \! \langle |P_k| \rangle \! \rangle$ that are time-averaged between
$100 \hbar/E_\text{F}$ and $400 \hbar/E_\text{F}$. As shown in Fig.~\ref{fig:PanelsTramp0}, for a sudden ramp,
the magnitudes of the pair amplitudes settle quickly as the snapshot distribution at $E_\text{F} t = 3 \hbar$ already
agrees approximately with the one obtained via time-averaging. However, even at long times, 
this distribution takes much larger values than the ones expected in equilibrium at the final interaction
strength $\frac{1}{k_\text{F} a} = -1.3493$. This result explains the large finite value of the sum over the magnitude of the pair
amplitudes presented in Fig.~\ref{fig:OPvsTime} signalling that Cooper pairs survive through the quench (even though with a smaller amplitude
than in the initial state).

As shown in the central panel of Fig.~\ref{fig:PanelsTramp0}, for the sudden ramp, each pair rapidly acquires a particular
phase proportional to $2 E_\tbf{k}$ leading to complete dephasing such that already at $E_\text{F} t = 10 \hbar$ the superconducting order
parameter is totally obliterated. 

The evolution of the phases can be understood via the Fourier transform of the pair amplitudes, $|\mathcal{F}[\text{Im}[P_\tbf{k}(t)]]|$,
and as $\text{Im}[P_\tbf{k}(t)]$ and $\text{Re}[P_\tbf{k}(t)]$ provide the same information about the phase evolution, we only
consider the former without loss of generality.
From this quantity, we see that the sudden ramp generates quasiparticle pairs at $-2 \mu_\text{f} \pm 2 E_\tbf{k}$
(see lower panel of Fig.~\ref{fig:PanelsTramp0}) with $\mu_\text{f}$ the chemical potential at the final interaction strength.
These quasiparticle pairs are at the origin of the parabolic distribution of the phases (central panel). This result indicates
that the system dynamically organizes into a non-thermal state made of pre-formed but dephased Cooper pairs.

The pair amplitude distributions are also non-trivially affected when the interaction strength is slowly ramped down.
For the ramp time $E_\text{F} t_\text{ramp} = 6 \hbar$, both the
snapshot and time-averaged distributions are clearly finite and non-thermal. Only the small and large momentum tails of the time-averaged distribution
agree with the thermal equilibrium distribution at $\frac{1}{k_\text{F} a} = -1.3493$. The temperature $T$ used in Fig.~\ref{fig:PanelsTramp6}
is found by solving the finite-temperature gap equation~\cite{Tinkham} assuming that
$\langle \! \langle |\Delta| \rangle \! \rangle = \Delta(T)$.

As we see in the lower panel of Fig.~\ref{fig:PanelsTramp6}, the $E_\text{F} t_\text{ramp} = 6 \hbar$ ramp creates fewer
quasiparticle pairs at  $-2 \mu_\text{f} \pm 2 E_\tbf{k}$ and all $P_\tbf{k}$ signals have a strong component at $-2\mu_\text{f}$. At short
times compared to the ramp duration, the phase remains fully locked, then as the evolution goes on, in the momentum interval where
most of the quasiparticle pairs are generated, each Cooper pair begins accumulating a particular phase. This process leads to a partial
loss of phase coherence, but, as shown in Fig.~\ref{fig:OPvsTime}, the Cooper pairs are still sufficiently synchronized for
superconductivity to survive at the considered times.


Finally, for $E_\text{F} t_\text{ramp} = 30 \hbar$, we find that the dynamics enters a different regime as the distribution of
pair amplitudes becomes thermal. As one sees from Fig.~\ref{fig:PanelsTramp30}, the distribution obtained at
the end of the ramp strongly resembles the one expected for a superconducting system in equilibrium at $T =  0.35 T_c$
for an interaction strength of $\frac{1}{k_\text{F} a} = -1.3493$. For this ramp schedule, quasiparticle pairs are solely generated
in a small momentum region around $k_\text{F}$ (see lower panel of Fig.~\ref{fig:PanelsTramp30}). The phase coherence remains
for the most part undisturbed by the interaction ramps. As illustrated in Fig.~\ref{fig:PanelsTramp30}, during the
ramp the phase starts to ripple around $k_\text{F}$, the region where quasiparticle pairs are generated, but phase locking
is for the most part maintained throughout the system.

%
%
To summarize, we analyzed the non-equilibrium dynamics of a BCS superconductor when the interaction strength is slowly ramped down.
We identified three different dynamical regimes and, in particular, we demonstrated the dynamical creation of a steady state of
pre-formed pairs without global phase coherence. The insights gained from this study will likely pave the way to employ slow quenches
to create other steady states with novel properties absent in thermal equilibrium.

{\it Acknowledgments:} We thank Kuiyi Gao for useful discussions.
We acknowledge funding from the European Research Council (ERC) under the Horizon
2020 research and innovation programme, grant agreement No. 648166 (Phonton) and No. 616082 (UpFermi),
and from the Deutsche Forschungsgemeinschaft (DFG, German Research Foundation) project number 277625399 - TRR 185
project B4 and project number 277146847 - project C05. 


\bibliography{jkombe_slow_quench_BCS}

\end{document}